\documentclass[aps,groupedaddress,floatfix]{revtex4}
\usepackage{amssymb}
\usepackage{amsmath}
\usepackage{graphicx}
\usepackage{color}
\begin{document}

\vspace{5mm}

\title{Self-averaging in the random 2D Ising ferromagnet}

\author{Victor Dotsenko$^{\, a,b}$, Yurij Holovatch$^{\, c,e}$, Maxym Dudka$^{\,
    c,e}$, and Martin Weigel$^{\,d,e}$}

\affiliation{$^a$LPTMC, Universit\'e Paris VI, 75252 Paris, France}

\affiliation{$^b$L.D.\ Landau Institute for Theoretical Physics,
   119334 Moscow, Russia}

\affiliation{$^c$Institute for Condensed Matter Physics, National Acad. Sci. of Ukraine,
 79011 Lviv, Ukraine}

\affiliation{$^d$Applied Mathematics Research Centre, Coventry University, Coventry, CV1 5FB, United Kingdom}

\affiliation{$^e$${\mathbb L}^4$ Collaboration \&
Doctoral College for the Statistical Physics of Complex Systems,
Leipzig-Lorraine-Lviv-Coventry, D-04009 Leipzig, Germany}

\date{\today}

\begin{abstract}
  We study sample-to-sample fluctuations in a critical two-dimensional Ising model
  with quenched random ferromagnetic couplings. Using replica calculations in the
  renormalization group framework we derive explicit expressions for the probability
  distribution function of the critical internal energy and for the specific heat
  fluctuations. It is shown that the disorder distribution of internal energies is
  Gaussian, and the typical sample-to-sample fluctuations as well as the average
  value scale with the system size $L$ like $\sim L \ln\ln(L)$. In contrast, the
  specific heat is shown to be self-averaging with a distribution function that tends
  to a $\delta$-peak in the thermodynamic limit $L \to \infty$.  While previously a
  lack of self-averaging was found for the free energy, we here obtain results for
  quantities that are directly measurable in simulations, and implications for
  measurements in the actual lattice system are discussed.
\end{abstract}

\maketitle

\medskip

\section{Introduction}

A varying degree of impurities is present in every material studied in the
laboratory. The consequences of disorder vary strongly from system to system,
however. While for strong disorder randomness is accompanied by frustration effects
and often leads to the absence of long-range order \cite{binderyoung}, the case of
weak disorder is less spectacular in that it cannot destroy the low temperature
ferromagnetic ground state \cite{Pelissetto_rev,Dotsenko_rev, Holovatch_rev}. Still,
in many cases one observes a change in the character of the transition to this
ferromagnetic phase. For pure systems with continuous phase transitions, as revealed
by Harris \cite{Harris74} weak disorder is relevant for the critical behavior only if
the specific heat is divergent, i.e., the corresponding critical exponent
$\alpha > 0$, as in these cases the random fluctuations grow faster with system size
than the energy fluctuations. The critical behavior is then governed by a new, random
renormalization-group fixed point, and the pure fixed point becomes unstable. On the
other hand, first-order phase transitions in pure systems are softened by the
addition of weak disorder and, in some cases, are turned into continuous transitions
\cite{first-to-second,cardy:99a}. These effects of weak disorder have been thoroughly
studied both analytically \cite{Harris74b,Khmelnitskii75,Lubensky75,Grinstein76} as
well as numerically \cite{landau:80,ballesteros:98,chatelain:01}, see
Ref.~\cite{Folk_Ising} for a review.

An intriguing aspect of systems with quenched disorder is related to the possibility
of exceedingly strong disorder induced fluctuations. In some cases, these might lead
to a loss of self-averaging
\cite{NonSelfAverage1,NonSelfAverage2,NonSelfAverage3,RandFerro}, i.e., the behavior
of a large sample with a specific realization of impurities such as an actual
material sample in the laboratory will not be well described by the ensemble average
normally calculated in an analytical or numerical approach. This clearly has profound
consequences for the physical interpretation of the outcomes and the possibilities
for comparing theoretical and experimental results. The presence or absence of
self-averaging is connected to the question of the relevance of disorder for the
system studied \cite{NonSelfAverage2,NonSelfAverage3}, and it affects static as well
as dynamic properties \cite{Tomita2001,janke:00}. Recently, an explicit expression
for the probability distribution function of the critical free-energy fluctuations
for a weakly disordered Ising ferromagnet was derived for $d<4$ and its universal
shape was obtained at $d=3$ \cite{Dotsenko14}. As free energies are not directly
accessible in experimental or numerical studies, however, it is desirable to study
the self-averaging properties of directly measurable quantities.

A system of particular interest is the Ising model in two
dimensions, where a wealth of exact results are available for the
pure case \cite{onsager}. When weak disorder in the form of random,
but non-frustrating bonds is added, the Harris criterion is unable
to decide its significance as $\alpha=0$ and the system hence
provides a marginal case. Still, it is now well established that
such weak disorder ``marginally'' modifies the critical behavior of
this system so that the logarithmic singularity of the specific heat
is changed into a double logarithmic one
\cite{DD,DDf,2DRIMa,2DRIMb}. While a number of further aspects of
this problem have been studied, such as the effect of correlated
disorder in the form of extended impurities \cite{correlated},  the
question of the disorder distribution of
  measurable quantities and their (lack of) self-averaging behavior was less studied.
  Following the seminal works \cite{NonSelfAverage1,NonSelfAverage2} the relative
  variance of thermodynamic observables was usually studied as a measure to gauge the
  presence or absence of self-averaging. It was shown that for irrelevant disorder
  the relative variance weakly decreases as a power of $L$ indicating the presence of
  ``weak self-averaging'', while for relevant disorder this ratio approaches a
  non-zero constant as $L\to\infty$, indicating a lack of self-averaging
  \cite{NonSelfAverage2}.  Results of numerical studies of this quantity for the
  disordered two-dimensional Ising model
  \cite{NonSelfAverage1,Tomita2001,2DRIMa,FytasMalakis10}, where the disorder is
  marginally relevant, were not completely conclusive. Here we derive the form of the
  distribution functions of sample-to-sample fluctuations and discuss their
  asymptotics as $L\to\infty$.

The rest of the paper is organized as follows. In Sec.~\ref{II} we recall the
description of the critical two-dimensional Ising model in terms of free Majorana
fermions and show how this description can be extended to the disordered system. In
Sec.~\ref{III} we introduce the replica formalism for the energy distribution of the
model. Section~\ref{IV} contains the renormalization group calculations for the
disorder distribution of the internal energy, where we show that the internal energy
lacks self averaging at criticality. The typical value of its sample-to-sample
fluctuations scale with the system size $L$ in the same way as its average
$\sim L \ln\ln(L)$. In Sec.~\ref{V} we extend this calculation to the specific heat
and see that in contrast the energy it is self-averaging, and its distribution turns
into a $\delta$-function in the limit $L \to \infty$. Finally, Sec.\ref{VI} contains
our conclusions.

\section{The model \label{II}}

It is well known that the critical behavior of the two-dimensional ferromagnetic
Ising model can be described in terms of free two-component Grassmann-Majorana spinor
fields $\psi({\bf r}) = \big(\psi_{1}({\bf r}), \, \psi_{2}({\bf r})\big)$ with the
following Hamiltonian (see e.g. \cite{Itzykson89}):
\begin{equation}
 \label{1}
 H_{0}[\psi; \tau] \; = \; \frac{1}{2} \int d^{2}r \, \bigl[
 \overline{\psi}({\bf r})\hat{\partial} \psi({\bf r}) \; + \; \tau \, \overline{\psi}({\bf r})\psi({\bf r})
 \bigr],
\end{equation}
where $\tau \propto (T-T_{c})/T_{c} \, \ll \, 1$ and $T_c$ denotes the critical
temperature [in what follows, to simplify formulas we define
$\tau \equiv (T-T_{c})/T_{c}$)]. Further,
\begin{equation}
 \label{2}
 \hat{\partial} \; = \; \hat{\sigma_{1}} \frac{\partial}{\partial x} \; + \;
                        \hat{\sigma_{2}} \frac{\partial}{\partial y},
\end{equation}
where
\begin{equation}
 \label{3}
\hat{\sigma_{1}} \; = \;
  \begin{pmatrix}
    0 & 1 \\
    1 & 0
  \end{pmatrix}, \; \; \; \;
  \hat{\sigma_{2}} \; = \;
  \begin{pmatrix}
    0 & -i \\
    i & 0
  \end{pmatrix}, \; \; \; \;
  \hat{\sigma_{3}} \; = \; \hat{\sigma_{1}}\, \hat{\sigma_{2}} \; = \;
  \begin{pmatrix}
    i & 0 \\
    0 & -i
  \end{pmatrix},
\end{equation}
are the Pauli matrices, and $\overline{\psi} \equiv \psi \, \hat{\sigma_{3}}$.
At a given value of the temperature parameter $\tau$ the partition function $Z(\tau)$
of the system (\ref{1}) is
\begin{equation}
 \label{6}
 Z(\tau) = \int {\cal D} \psi \; \exp\bigl\{-H_{0}[\psi; \tau]\bigr\},
\end{equation}
where the integration measure is defined as
\begin{equation}
 \label{8}
 \int {\cal D} \psi = \prod_{{\bf r}} \; \Biggl[ - \int \, d\psi_{1}({\bf r}) \, d\psi_{2}({\bf r}) \Biggr],
\end{equation}
and the integration and commutation rules are
\begin{equation}
 \label{9}
 \int \, d\psi_{\alpha}({\bf r}) \; = \; 0 \; ,
 \; \; \;
 \int \, d\psi_{\alpha}({\bf r}) \, \psi_{\alpha}({\bf r}) \; = \;
 -\int \, \psi_{\alpha}({\bf r}) \, d \psi_{\alpha}({\bf r}) \; = \; 1,
\end{equation}
\begin{equation}
 \label{10}
 \psi_{\alpha}({\bf r}) \psi_{\beta}({\bf r}') \; = \; - \psi_{\beta}({\bf r}') \psi_{\alpha}({\bf r}) \; , \; \; \;
 \left[\psi_{\alpha}({\bf r}) \right]^{2} \; = \; 0 \, .
\end{equation}
Hence the free energy is
\begin{equation}
 \label{7}
 F(\tau) \; = \; -\ln \bigl[ Z(\tau)\bigr].
\end{equation}
Note that we did not include the usual temperature prefactor in the definition of the
free energy (\ref{7}). Our analysis is performed close to $T_{c}$, which for
simplicity is taken to be 1, and we are looking {\em only} for the leading terms
(singularities) in the parameter $\tau = (T-T_{c})/T_{c} \, = \, T -
1$. Therefore, in the limit $\tau \to 0$,
\begin{equation}
 \label{d1a}
 F(\tau) \; = \; - T \, \ln \bigl[ Z(\tau)\bigr] \, = -  \, \ln \bigl[ Z(\tau)\bigr] - \, \tau \, \ln \bigl[ Z(\tau)\bigr] \, = \,
 -  \, \ln \bigl[ Z(\tau)\bigr] \; + \; O(\tau) \, .
\end{equation}
Simple integration of Eq.~(\ref{6}) yields
\begin{equation}
 \label{11}
 Z(\tau) \; = \; \Bigl[\det\big(\hat{\partial} + \tau \hat{\sigma_{0}}\bigr)\Bigr]^{1/2},
\end{equation}
where $\hat{\sigma_{0}}$ is the unit matrix, and the term on the right hand side is a
symbolic notation for the determinant of the $L^{2} \times L^{2}$ matrix defining the
Hamiltonian (\ref{1}) written in a discrete way on an $L\times L$ lattice. The free
energy reads
\begin{equation}
 \label{12}
 F(\tau) \; = \; - \frac{1}{2} \ln\Bigl[\det\big(\hat{\partial} + \tau \hat{\sigma_{0}}\bigr)\Bigr] \; \sim \;
 - L^{2} \int_{|p|< 1} d^{2} p \; \ln\bigl(p^{2} + \tau^{2}\bigr).
\end{equation}
Note that the celebrated logarithmic divergence of the specific heat in the limit
$\tau \to 0$ follows immediately from Eq. (\ref{12}):
\begin{equation}
 \label{13}
 C(\tau) \; = \; - \frac{\partial^{2}}{\partial \tau^{2}} F(\tau) \; \sim \;
 L^{2} \int_{|p|< 1} \frac{d^{2} p}{p^{2} + \tau^{2}} \; \sim \;
 L^{2} \int_{|\tau|}^{1} \frac{d p}{p} \; \sim \;
 L^{2} \ln\frac{1}{|\tau|}.
\end{equation}

The presence of weak quenched disorder in the considered system can
be described by allowing for a spatially varying local transition
temperature $T_{c}$ which, in turn, can be represented by quenched
spatial fluctuations of the temperature parameter $\tau$ in the
Hamiltonian (\ref{1}) (see, e.g., Ref.~\cite{DD}).  In other words,
the critical behavior of the weakly disordered two-dimensional Ising
model can be described by the spinor Hamiltonian
\begin{equation}
 \label{14}
 H[\psi; \tau, \delta\tau] \; = \; \frac{1}{2} \int d^{2}r \, \Bigl[
 \overline{\psi}({\bf r})\hat{\partial} \psi({\bf r}) \; + \;
 \bigl(\tau + \delta\tau({\bf r})\bigr)\, \overline{\psi}({\bf r})\psi({\bf r})
 \Bigr],
\end{equation}
where the random function $\delta\tau({\bf r})$ is characterized as a spatially
uncorrelated Gaussian distribution with zero mean,
$\overline{\delta\tau({\bf r})} = 0$, and variance
\begin{equation}
\label{15}
\overline{\delta\tau({\bf r})\delta\tau({\bf r}')} \; = \; 2 g_{0} \, \delta({\bf r} - {\bf r}'),
\end{equation}
where the parameter $g_{0} \ll 1$ defines the disorder strength.
For a given realization of the quenched function $\delta\tau({\bf r})$ the partition function of the
considered system is
\begin{equation}
 \label{16}
 Z[\tau; \delta\tau] \; = \; \int {\cal D} \psi \; \exp\bigl\{-H[\psi; \tau, \delta\tau ]\bigr\} \; = \;
 \exp\bigl\{ - F[\tau; \delta\tau]\bigr\},
\end{equation}
where $F[\tau; \delta\tau]$ is a random free-energy function. The internal energy of
a given realization is the first derivative of this free energy with respect to the
temperature parameter:
\begin{equation}
 \label{17}
 E[\tau; \delta\tau] \; = \; \frac{\partial}{\partial\tau} F[\tau; \delta\tau].
\end{equation}
It is clear that $E[\tau; \delta\tau]$ must be a singular function of $\tau$ in the
limit $\tau \to 0$ (in the pure system $E_{0}(\tau) \sim \tau \ln(1/|\tau|)$).
Additionally, $E[\tau; \delta\tau]$ also must be a {\it random\/} function exhibiting
sample-to-sample fluctuations.  The distribution function of these fluctuations is
main target of the present study.

\section{Replica formalism  \label{III}}

>From the definition (\ref{17}) we have
\begin{equation}
 \label{18}
 E[\tau; \delta\tau] \; = \; \lim_{\epsilon \to 0} \;
 \frac{1}{\epsilon}\bigl(F[\tau + \epsilon; \delta\tau] \; - \; F[\tau; \delta\tau]\bigr).
\end{equation}
Thus for a given finite value of $\epsilon$ (which has to be sent to zero at the end) we have
\begin{equation}
 \label{19}
 \epsilon \, E[\tau; \delta\tau] \; = \;
F[\tau + \epsilon; \delta\tau] \; - \; F[\tau; \delta\tau].
\end{equation}
According to the definition of the free energy, Eq. (\ref{16}), the above relation
can be represented in terms of the ratio of two partition functions,
\begin{equation}
 \label{20}
 \exp\bigl\{-\epsilon \, E[\tau; \delta\tau]\bigr\} \; = \;
 Z[\tau+\epsilon; \delta\tau] \, Z^{-1}[\tau; \delta\tau].
\end{equation}
Taking the $N$th power of both sides of the above equation and performing the
disorder average we find
\begin{equation}
 \label{21}
\int \, dE \; P_{\tau} (E) \; \exp(-\epsilon N E) \; = \;
 \overline{Z^{N}[\tau+\epsilon; \delta\tau] \, Z^{-N}[\tau; \delta\tau]}.
\end{equation}
Here, $P_{\tau} (E)$ is the probability distribution over disorder of the internal
energy of the system at a given value of the temperature parameter $\tau$ and
$\overline{(...)}$ denotes the average over the random functions
$\delta\tau({\bf r})$. Following the standard tricks of the replica formalism the
above relation can be represented in the following way:
\begin{equation}
 \label{22}
\int \, dE \; P_{\tau} (E) \; \exp(-\epsilon N E) \; = \; \lim_{M\to 0}
 \overline{Z^{N}[\tau+\epsilon; \delta\tau] \, Z^{M-N}[\tau; \delta\tau]} \; \equiv \;
 \lim_{M\to 0} \; {\cal Z}(M, N; \tau, \epsilon).
\end{equation}
In terms of this formalism, first it is assumed that both $M$ and $N$ are integers
such that $M > N$. Then, after deriving ${\cal Z}(M, N; \tau, \epsilon)$ as an
analytic function of $M$ and $N$, these parameters are analytically continued to
arbitrary real values and the limit $M\to 0$ is taken. Finally, we introduce a new
analytic parameter $s = \epsilon N$ and, provided that it exists, take the limit
$\epsilon \to 0$, such that the relation (\ref{22}) becomes the Laplace transform of
the probability distribution function $P_{\tau} (E)$,
\begin{equation}
 \label{23}
\int \, dE \; P_{\tau} (E) \; \exp(-s \, E) \; = \;
\lim_{\epsilon\to 0} \lim_{M\to 0} \;  {\cal Z}(M, s/\epsilon; \tau, \epsilon) \; \equiv \;
\tilde{{\cal Z}} (s, \tau).
\end{equation}
Thus, the above procedure, although it is not well-founded from a mathematical point
of view, at least formally allows to reconstruct the function $P_{\tau} (E)$ by the
inverse Laplace transform:
\begin{equation}
 \label{24}
 P_{\tau} (E) \; = \; \int_{-i\infty}^{+i\infty}\frac{ds}{2\pi i} \; \tilde{{\cal Z}}
 (s, \tau) \exp( s E ).
\end{equation}

To proceed, consider the structure of the replica partition function
${\cal Z}(M, N; \tau, \epsilon)$. According to the definitions (\ref{16}) and
(\ref{22}),
\begin{equation}
 \label{25}
 {\cal Z}(M, N; \tau, \epsilon) \; = \;
 \int {\cal D}\psi \; \overline{\Biggl(\exp\Bigl\{
 -\sum_{a=1}^{N} H[\psi_{a}; \tau+\epsilon, \delta\tau] \; -
 \sum_{a=N+1}^{M} H[\psi_{a}; \tau, \delta\tau]
 \Bigr\}\Biggr)}.
\end{equation}
Substituting here the Hamiltonian (\ref{14}) and performing Gaussian averaging over
$\delta\tau({\bf r})$ using Eq. (\ref{15}) we find:
\begin{equation}
 \label{26}
 {\cal Z}(M, N; \tau, \epsilon) \; = \;  \int {\cal D}\psi \; \exp\Bigl\{- {\cal H}_{M,N}[\psi; \tau, \epsilon]\Bigr\}
 \; \equiv \; \exp\bigl\{- {\cal F}(M,N;\tau,\epsilon)\bigr\},
\end{equation}
where ${\cal F}(M,N;\tau,\epsilon)$ can be called the ``replica free energy'' and
\begin{equation}
 \label{27}
 {\cal H}_{M,N}[\psi; \tau, \epsilon] \; = \;  \int d^{2}r \, \Biggl[
 \frac{1}{2}\sum_{a=1}^{M}\overline{\psi}_{a}({\bf r})\hat{\partial} \psi_{a}({\bf r}) \; + \;
 \frac{1}{2}\sum_{a=1}^{M} m_{a} \bigl(\overline{\psi}_{a}({\bf r})\psi_{a}({\bf r})\bigr) \; - \;
 \frac{1}{4} g_{0} \sum_{a,b=1}^{M} \bigl(\overline{\psi}_{a}({\bf r})\psi_{a}({\bf r})\bigr)
                                    \bigl(\overline{\psi}_{b}({\bf r})\psi_{b}({\bf r})\bigr)
 \Biggr],
\end{equation}
where
\begin{equation}
 \label{28}
 m_{a} \; = \; \left\{
                          \begin{array}{ll}
(\tau+\epsilon)\, \; \; \mbox{for} \; a = 1, ..., N \, ,
\\
\\
\tau \, \; \; \; \; \; \;  \; \; \; \; \, \mbox{for} \; a = N+1,
..., M \, .
                          \end{array}
\right.
\end{equation}
The expression obtained, Eq.~(\ref{27}), has the form of an effective Hamiltonian of
the random Ising model, but with replica-dependent masses. As we will see below, this
difference will further influence properties of the internal energy distribution.  In
the next section we will derive the function ${\cal F}(M, N; \tau, \epsilon)$ of
Eq.~(\ref{26}) using standard procedures of the renormalization group approach.

\section{Renormalization group calculations  \label{IV}}

It is well known that the spinor-field theory with four-fermion interactions is
renormalizable in two dimensions, and the renormalization equations lead to
``zero-charge'' asymptotics for the charge $g$ and mass $m$ (see, e.g.,
Ref.~\cite{DD}). Renormalization of the replica Hamiltonian (\ref{27}) can be
achieved in a standard way by integrating out short wave-length degrees of freedom in
the band $\tilde{\Lambda} < p < \Lambda$, where $\Lambda$ and $\tilde{\Lambda}$ are
the old and new ultraviolet momentum cut-offs, respectively. One can easily show that
the renormalization of the charge $g$ and the mass $m_{a}$ in the Hamiltonian
(\ref{27}) is given by the following equations [cf.~Eqs. (4.28) in Ref.~\cite{DD}]:
\begin{eqnarray}
 \label{29}
 \frac{d}{d\xi} g(\xi) &=& -\frac{1}{\pi}(2 - M) \; g^{2}(\xi),
 \\
 \nonumber
 \\
 \frac{d}{d\xi} m_{a}(\xi) &=& -\frac{1}{\pi}\Bigl(m_{a}(\xi) \; - \; \sum_{b=1}^{M} m_{b}(\xi)\Bigr) \; g(\xi),
 \label{30}
\end{eqnarray}
where $\xi = \ln\bigl(\Lambda/\tilde{\Lambda}\bigr)$ and
\begin{equation}
 \label{31}
 m_{a}(\xi) \; = \; \left\{
                          \begin{array}{ll}
\tilde{m}(\xi)\,
\; \;
\mbox{for} \; a = 1, ..., N
\\
\\
m(\xi) \,
\; \;
\mbox{for} \; a = N+1, ..., M,
                          \end{array}
\right.
\end{equation}
with the initial conditions $g(0) =  g_{0}$, $\; \tilde{m}(0) =
(\tau+\epsilon)$ and $m(0) = \tau$. Substituting Eq. (\ref{31}) into
Eq. (\ref{30}) we get:
\begin{eqnarray}
 \label{32}
 \frac{d}{d\xi} \tilde{m}(\xi) &=& -\frac{1}{\pi}\Bigl[\tilde{m}(\xi) \; - \; N \tilde{m}(\xi) - (M-N) m(\xi)\Bigr] \; g(\xi),
 \\
 \nonumber
 \\
 \frac{d}{d\xi} m(\xi) &=& -\frac{1}{\pi}\Bigl[m(\xi) \; - \; N \tilde{m}(\xi) - (M-N) m(\xi)\Bigr] \; g(\xi).
\label{33}
\end{eqnarray}
The solution of Eq.~(\ref{29}) is
\begin{equation}
 \label{34}
 g(\xi) \; = \; \frac{g_{0}}{1 \; + \; \frac{1}{\pi}(2-M) \, g_{0} \,
 \xi} \, .
\end{equation}
Equations (\ref{29})--(\ref{30}) have been obtained in the one loop
approximation. The two-loop approximation has been studied in Ref.~\cite{Ludwig87},
where it was shown that it gives only a next-order logarithmic correction to the
one-loop result. Therefore, being interested in the leading asymptotics, we proceed
further within the one-loop approximation. Substituting the above solution (\ref{34})
into Eqs.~(\ref{32})--(\ref{33}) in the limit $M\to 0$ one easily finds \footnote{
  One can easily check that in both cases: (i) solving these equations for a given
  finite value of $M$ and only after that putting $M=0$, or (ii) putting $M=0$ right
  away in the Eqs.~(\ref{32})--(\ref{33}) and after that solving them, one obtains
  the same result.}:
\begin{eqnarray}
 \label{35}
 m(\xi) &=& \Biggl[\tau \; + \; \frac{1}{2} (\epsilon N) \, \ln\Bigl(1 \; + \; \frac{2}{\pi}\, g_{0} \, \xi\Bigr)\Biggr] \,
            \Delta(\xi),
 \\
 \nonumber
 \\
  \tilde{m}(\xi) &=& m(\xi) \; + \; \epsilon \, \Delta(\xi),
  \label{36}
  \\
 \nonumber
 \\
 \Delta(\xi) &=& \frac{1}{\sqrt{1 \; + \; \frac{2}{\pi}\, g_{0} \, \xi}}.
 \label{37}
\end{eqnarray}

The critical properties of a model with ``zero-charge''
renormalization [according to Eq. (\ref{34}), $g(\xi\to\infty) \sim
1/\xi \to 0$] can be studied exactly by renormalization group
methods \cite{Larkin-Khmelnitskii,Aharony} (see also \cite{DD}).
According to the standard procedure of RG calculations, the
  singular contribution of thermodynamic quantities in the vicinity of the critical
  point is obtained by using only the non-interacting part of the renormalized
  Hamiltonian [the first two terms of the Hamiltonian (\ref{27})], in which the mass
  terms $m_{a}$, $a=1$, $\ldots$, $M$, become scale-dependent parameters,
  Eqs.~(\ref{31}) and (\ref{35})--(\ref{37}).  In other words, in the process of the
  RG procedure the contributions originating in the interaction terms of the
  Hamiltonian (\ref{26}) are effectively ``absorbed'' into the mass terms. In this
  case, similar to the pure system [see Eqs.~(\ref{11})--(\ref{12})], we get:
\begin{eqnarray}
 \nonumber
 {\cal F}(0,N;\tau,\epsilon) &=& - \lim_{M\to 0}\ln\Bigl[{\cal Z}(M,N;\tau,\epsilon)\Bigr]
 \\
 \nonumber
 \\
\nonumber
 &=& - L^{2} \lim_{M\to 0} \int_{|p|<1} \frac{d^{2}p}{(2\pi)^{2}} \;
 \ln\Biggl[\prod_{a=1}^{M} \,\det\Bigl(i\hat{p} + m_{a}(p)\hat{\sigma}_{0}\Bigr)^{1/2} \Biggr]
\\
 \nonumber
 \\
\nonumber
 &=& - L^{2} \lim_{M\to 0} \int_{|p|<1} \frac{d^{2}p}{(2\pi)^{2}} \;
 \ln\Biggl[
 \det\Bigl(i\hat{p} + \tilde{m}(p)\hat{\sigma}_{0}\Bigr)^{\frac{N}{2}} \times
 \det\Bigl(i\hat{p} + m(p)\hat{\sigma}_{0}\Bigr)^{\frac{M-N}{2}}
 \Biggr],
\\
 \nonumber
 \\
 &=& - L^{2} \int_{|p|<1} \frac{d^{2}p}{(2\pi)^{2}} \;
 \ln\Biggl[
 \det\Bigl(i\hat{p} + \tilde{m}(p)\hat{\sigma}_{0}\Bigr)^{N/2} \times
 \det\Bigl(i\hat{p} + m(p)\hat{\sigma}_{0}\Bigr)^{-N/2}
 \Biggr],
\label{38}
\end{eqnarray}
where [cf.\ Eqs.~(\ref{2})--(\ref{3})]
\begin{equation}
 \label{39}
 \hat{p} \; = \; \hat{\sigma}_{1} p_{x} \; + \; \hat{\sigma}_{2} p_{y}.
\end{equation}
Here the mass parameters $m(p)$ and $\tilde{m}(p)$ are taken to be dependent on the
scale according to Eqs.~(\ref{35})--(\ref{37}) with $\xi = \ln(1/p)$. Simple
calculations yield [cf.~  Eq.~(11)]
\begin{eqnarray}
 \nonumber
 {\cal F}(0,N;\tau,\epsilon) &=& - L^{2} \int_{|p|<1} \frac{d^{2}p}{(2\pi)^{2}} \;
 \Biggl[
 \frac{1}{2} N \ln\bigl(p^{2} + \tilde{m}^{2}(p)\bigr) - \frac{1}{2} N \ln\bigl(p^{2} + m^{2}(p)\bigr)
 \Biggr]
 \\
 \nonumber
 \\
 &=& -\frac{1}{4\pi} L^{2} N \int_{0}^{1} dp \, p \;
 \ln\Biggl[
 \frac{p^{2} + \bigl(m(p) + \epsilon \Delta(p)\bigr)^{2}}{p^{2} + m^{2}(p)}
 \Biggr].
 \label{40}
\end{eqnarray}
Substituting here the solutions (\ref{35})--(\ref{37}) in the leading order in
$\epsilon \to 0$ we get (see the Appendix for details):
\begin{eqnarray}
 \nonumber
 {\cal F}(0,N;\tau,\epsilon) &\simeq& -\frac{1}{4\pi} L^{2} N \int_{0}^{1} dp \, p \;
 \ln\Biggl[
 1 \; + \; \epsilon \, \frac{2 m(p) \Delta(p)}{p^{2} + m^{2}(p)}
 \Biggr]
 \\
 \nonumber
 \\
 \nonumber
 &\simeq& -\frac{1}{2\pi} L^{2} (\epsilon N) \int_{0}^{1} dp \, p \;
 \frac{ \tau + \frac{1}{2} (\epsilon N) \ln\bigl[1 + \frac{2}{\pi}g_{0}\ln(1/p)\bigr]}{
 \bigl[p^{2} + m^{2}(p)\bigr] \bigl[1 + \frac{2}{\pi}g_{0}\ln(1/p)\bigr]}
 \\
 \nonumber
 \\ \nonumber
 &\simeq&
 -\frac{1}{2\pi} L^{2} (\epsilon N) \int_{|\tau|}^{1} \frac{dp}{p} \;
 \frac{ \tau + \frac{1}{2} (\epsilon N) \ln\bigl[1 + \frac{2}{\pi}g_{0}\ln(1/p)\bigr]}{
 1 + \frac{2}{\pi}g_{0}\ln(1/p)}
 \\
 \nonumber
 \\
 &=&  E(\tau) (\epsilon N) \; - \; \frac{1}{2} E_{*}^{2}(\tau) (\epsilon N)^{2},
\label{41}
\end{eqnarray}
where
\begin{eqnarray}
\label{42}
E(\tau) &=& -\frac{1}{4 g_{0}} L^{2} \, \tau \, \ln\Bigl(1 + \frac{2}{\pi}g_{0}\ln(1/|\tau|)\Bigr),
\\
\nonumber
\\
E_{*}(\tau) &=& \frac{1}{2\sqrt{2 g_{0}}} L\ln\Bigl(1 + \frac{2}{\pi}g_{0}\ln(1/|\tau|)\Bigr).
\label{43}
\end{eqnarray}
  It should be stressed that the expression for the free energy
  ${\cal F}(0,N; \tau,\epsilon)$ obtained in Eq.~(\ref{41}) contains no higher-order
  terms in powers of $\epsilon N$ that were neglected.  On the other hand, the result
  is not exact and both $E(\tau)$ and $E_{*}(\tau)$ contain higher order logarithmic
  corrections of the form $\ln\ln(1/\tau)$ which were neglected in the limit
  $\tau \ll 1$ considered here [see the calculations given in the Appendix as well as
  the remark below Eq.~(\ref{34})].
Thus, for the replica partition function on the r.h.s.\ of relation (\ref{23}) for
the Laplace transform we obtain according to definition (\ref{26}):
\begin{equation}
 \label{44}
 \tilde{{\cal Z}} (s, \tau) \; = \; \lim_{\epsilon\to 0} \, \exp\bigl\{- {\cal F}(0,s/\epsilon;\tau,\epsilon)\bigr\}
 \; = \; \exp\Bigl\{ - E(\tau) \, s \; + \; \frac{1}{2} E_{*}^{2}(\tau) \, s^{2}\Bigr\}.
\end{equation}
Substituting this into the inverse Laplace transform relation (\ref{24}) we get:
\begin{equation}
  \label{45}
  P_{\tau} (E) \; = \; \int_{-i\infty}^{+i\infty}\frac{ds}{2\pi i} \;
 \exp\Bigl\{ - E(\tau) \, s \; + \; \frac{1}{2} E_{*}^{2}(\tau) \, s^{2} \; + \; s \, E \Bigr\},
\end{equation}
or
\begin{equation}
  \label{46}
  P_{\tau} (E) \; = \;\frac{1}{\sqrt{2\pi} E_{*}(\tau)} \,
  \exp\Biggl\{-\frac{\bigl(E - E(\tau)\bigr)^{2}}{2 E_{*}^{2}(\tau)}\Biggr\}.
\end{equation}
Thus, the sample-to-sample fluctuations of the critical internal energy of the weakly
disordered two-dimensional Ising model are described by a Gaussian distribution
characterized by the mean value $E(\tau)$ given in Eq.~(\ref{42}) and typical
deviations $E_{*}(\tau)$ as given in Eq.~(\ref{43}). Let us check the behavior of
$ P_{\tau} (E)$ as $L\to\infty$ and for $\tau\to 0$. If, in this limit the
distribution (\ref{46}) tends towards a $\delta$-function, then $E$ is
self-averaging, otherwise it is not.  For a fixed value of $\tau$
Eqs.~(\ref{42}),(\ref{43}) and (\ref{46}) reveal that the distribution function of
the energy density $e \equiv E/L^{2}$ in the thermodynamic limit turns into a
delta-function: $P(e) \, = \delta(e - e_{0}(\tau))$ with
$e_{0}(\tau) \, = \, -(\tau/4 g_{0})\, \ln\Bigl(1 +
\frac{2}{\pi}g_{0}\ln(1/|\tau|)\Bigr)$.
In other words, in this case the energy density is self-averaging.  On the other
hand, for finite $L$ the limit $\tau\to 0$ cannot be used directly in formulas
(\ref{42}), (\ref{43}) and (\ref{46}), since in this case the correlation length
($R_c(\tau)\sim 1/\tau$) exceeds the system size, which makes no physical sense. The
point is that the RG procedure must be stopped at scales of the order of the system
size $L$ (provided we take $\tau\ll L$ in the starting Hamiltonian). Therefore, one
will get the result given in Eqs. (\ref{42}), (\ref{43}) and (\ref{46}) again, but
there the parameter $\tau$ has to be replaced by $1/L$ which makes it temperature
independent as it should be.

In close vicinity of the critical point where the disorder induced critical behavior
sets in, i.e., at $\tau \ll \exp(-\pi/2g_{0})$, one finds:
\begin{eqnarray}
\label{47}
E(\tau) &\simeq& -\frac{1}{4 g_{0}} L^{2} \, \tau \, \ln\ln(1/|\tau|),
\\
\nonumber
\\
E_{*}(\tau) &\simeq& \frac{1}{2\sqrt{2 g_{0}}} L\ln\ln(1/|\tau|).
\label{48}
\end{eqnarray}
At large but finite value of the system size $L$, we expect the pseudo-critical
temperature to scale as $\tau_{c} \sim L^{-\nu} =  1/L$, and hence
\begin{eqnarray}
\label{49}
E_{c}(L) \equiv E(\tau=1/L) &\sim& -\frac{1}{g_{0}} L \, \ln\ln(L),
\\
\nonumber
\\
E^{*}_{c}(L) \equiv E_{*}(\tau = 1/L) &\sim& \frac{1}{\sqrt{g_{0}}} L \, \ln\ln(L).
\label{50}
\end{eqnarray}
Comparing Eqs.~(\ref{47}), (\ref{49}), (\ref{50}), and (\ref{46}) one concludes that
at sufficiently large system size, $L \; \gg \; \exp(\pi/2g_{0})$, the critical
internal energy $E$ can be written as a sum of its mean value and a fluctuating part:
\begin{equation}
 \label{52}
 E  \; \sim \; -\frac{1}{g_{0}} L \, \ln\ln(L) \; + \; \frac{1}{\sqrt{g_{0}}} L \, \ln\ln(L) \cdot  f,
\end{equation}
where the random quantity $f$ does not scale with $L$, $\; f \sim 1$, and is
described by a standard normal distribution
\begin{equation}
 \label{53}
 P_{c}(f) \; = \; \frac{1}{\sqrt{2\pi}} \, \exp\left(-\frac{1}{2} f^{2}\right).
\end{equation}
Eqs.~(\ref{52})--(\ref{53}) demonstrate that at criticality the internal energy of the
2D random-bond Ising ferromagnet is not self-averaging as the typical value of the
sample-to-sample fluctuations, $E_{c}^{*}(L) \sim g_0^{-1/2} L \ln\ln(L)$, scale with
the system size in the same way as its average value,
$E_{c}(L) \sim g_0^{-1} L \ln\ln(L)$. We note that the renormalization group
framework we use in our analysis gives access to the singular part of thermodynamic
functions only, and is not able to say anything about the behavior of non-singular
background terms that are present in a specific lattice realization. Therefore, it is
this singular part of the internal energy that is governed by the distribution
(\ref{46}).

\section{Specific Heat  \label{V}}

We now turn to an investigation of the behavior of the specific heat. To this end, we
repeat the steps performed in Secs.~\ref{II} and \ref{III} for the {\em second\/}
derivative of the free energy,
\begin{equation}
\label{54}
C[\tau; \delta \tau] \; = \; -\frac{\partial^{2}}{\partial \tau^{2}} F[\tau; \delta \tau].
\end{equation}
In terms of the replica formalism, instead of Eqs.~(\ref{18})--(\ref{22}) we get
\begin{equation}
 \label{55}
 C[\tau; \delta\tau] \; = \; -\lim_{\epsilon \to 0} \;
 \frac{1}{\epsilon^{2}}\bigl(F[\tau + \epsilon; \delta\tau] \; + \;
                             F[\tau - \epsilon; \delta\tau] \; - \;
                             2F[\tau; \delta\tau] \bigr),
\end{equation}
so that
\begin{equation}
 \label{56}
 \exp\bigl\{\epsilon^{2} \, C[\tau; \delta\tau]\bigr\} \; = \;
 Z[\tau+\epsilon; \delta\tau] \, Z[\tau-\epsilon; \delta\tau] \, Z^{-2}[\tau; \delta\tau]
\end{equation}
and
\begin{equation}
 \label{57}
\int \, dC \; {\cal P}_{\tau} (C) \; \exp\bigl(\epsilon^{2} N \, C\bigr) \; = \; \lim_{M\to 0}
 \overline{Z^{N}[\tau+\epsilon; \delta\tau] \,
           Z^{N}[\tau-\epsilon; \delta\tau] \,
           Z^{M-2N}[\tau; \delta\tau]} \; \equiv \;
 \lim_{M\to 0} \; {\cal Z}_{c}(M, N; \tau, \epsilon),
\end{equation}
where ${\cal P}_{\tau} (C)$ is the probability distribution function
of the specific heat. Correspondingly, instead of Eqs.
(\ref{23})--(\ref{24}) we have
\begin{equation}
 \label{58}
\int \, dC \; {\cal P}_{\tau} (C) \; \exp\bigl(s \, C\bigr) \; = \;
\lim_{\epsilon\to 0} \lim_{M\to 0} \;  {\cal Z}_{c}(M, s/\epsilon^{2}; \tau, \epsilon) \; \equiv \;
\tilde{{\cal Z}_{c}} (s, \tau),
\end{equation}
with $s=\epsilon^2 N$ in this case and
\begin{equation}
 \label{59}
 {\cal P}_{\tau} (C) \; = \; \int_{-i\infty}^{+i\infty}\frac{ds}{2\pi i} \; \tilde{{\cal Z}_{c}} (s, \tau) \exp\bigl( -s \, C \bigr),
\end{equation}
where
\begin{equation}
 \label{60}
 {\cal Z}_{c}(M, N; \tau, \epsilon) \; = \;  \int {\cal D}\psi \;
                 \exp\Bigl\{- {\cal H}^{(c)}_{M,N}[\psi; \tau, \epsilon]\Bigr\}
 \; \equiv \; \exp\bigl\{- {\cal F}_{c}(M,N;\tau,\epsilon)\bigr\},
\end{equation}
and the replica Hamiltonian ${\cal H}^{(c)}_{M,N}[\psi; \tau, \epsilon]$ is defined
by the r.h.s of Eq.~(\ref{27}) with
\begin{equation}
 \label{61}
 m_{a} \; = \; \left\{
                          \begin{array}{ll}
(\tau+\epsilon)\, \; \; \mbox{for} \; a = 1, ..., N,
\\
\\
(\tau-\epsilon)\, \; \; \mbox{for} \; a = N+1, ..., 2N,
\\
\\
\tau \, \; \; \; \; \; \;  \; \; \; \; \, \mbox{for} \; a = 2N+1,
..., M.
                          \end{array}
\right.
\end{equation}
The renormalization of the charge $g$ and the mass $m_{a}$ of this Hamiltonian is
given in Eqs.~(\ref{29})--(\ref{30}) where
\begin{equation}
 \label{62}
 m_{a} \; = \; \left\{
                          \begin{array}{ll}
m_{1}(\xi) \, \; \; \mbox{for} \; a = 1, ..., N,
\\
\\
m_{2}(\xi) \, \; \; \mbox{for} \; a = N+1, ..., 2N,
\\
\\
m(\xi) \, \; \; \; \; \; \;  \; \; \; \; \, \mbox{for} \; a = 2N+1,
..., M,
                          \end{array}
\right.
\end{equation}
with the initial conditions $m_{1}(0) = (\tau+\epsilon)$, $m_{2}(0)
= (\tau-\epsilon)$ and $m(0) = \tau$. One can easily show that in
the limit $M \to 0$, the sum
\begin{equation}
\label{63}
 \lim_{M\to 0}\left(\sum_{a=1}^{M} m_{a}(\xi) \right)\; =
\; N\Bigl(m_{1}(\xi) + m_{2}(\xi) - 2 m(\xi) \Bigr) \; \equiv \; 0,
\end{equation}
so that the solutions of the RG equations (\ref{29})--(\ref{30}) for
the masses $m_{1}(\xi)$, $m_{2}(\xi)$ and $m(\xi)$ turn out to be
effectively decoupled [unlike the situation for the internal energy,
Eqs.~(\ref{35})--(\ref{37})]:
\begin{eqnarray}
 \label{64}
 m_{1}(\xi) &=& \frac{\tau + \epsilon}{\sqrt{1 \; + \; \frac{2}{\pi}\, g_{0} \, \xi}},
 \\
 \nonumber
 \\
 m_{2}(\xi) &=& \frac{\tau - \epsilon}{\sqrt{1 \; + \; \frac{2}{\pi}\, g_{0} \, \xi}},
  \label{65}
  \\
 \nonumber
 \\
 m(\xi) &=& \frac{\tau}{\sqrt{1 \; + \; \frac{2}{\pi}\, g_{0} \, \xi}}.
 \label{66}
\end{eqnarray}
Correspondingly, instead of Eq.~(\ref{40}) we obtain
\begin{eqnarray}
 \nonumber
 {\cal F}_{c}(0,N;\tau,\epsilon) &=& - L^{2} \int_{|p|<1} \frac{d^{2}p}{(2\pi)^{2}} \;
 \Biggl[
  \frac{1}{2} N \ln\bigl(p^{2} + m_{1}^{2}(p)\bigr)
+ \frac{1}{2} N \ln\bigl(p^{2} + m_{2}^{2}(p)\bigr)
-             N \ln\bigl(p^{2} + m^{2}(p)\bigr)
 \Biggr]
 \\
 \nonumber
 \\
 &=& -\frac{1}{4\pi} L^{2} N \int_{0}^{1} dp \, p \;
 \ln\Biggl[
 \frac{\bigl(p^{2} + m_{1}^{2}(p)\bigr) \bigl(p^{2} + m_{2}^{2}(p)\bigr)}{\bigl(p^{2} + m^{2}(p)\bigr)^{2}}
 \Biggr].
 \label{67}
\end{eqnarray}
Substituting here the solutions (\ref{64})--(\ref{66}) in the leading
order in $\epsilon \to 0$ we get [c.f. Eq. (\ref{40})]
\begin{eqnarray}
 \nonumber
 {\cal F}_{c}(0,N;\tau,\epsilon) &\simeq& -\frac{1}{\pi} L^{2} \epsilon^{2} N
 \int_{0}^{1} \frac{dp \, p}{
 \bigl(p^{2} + m^{2}(p)\bigr) \Bigl(1 \; + \; \frac{2}{\pi}\, g_{0} \, \ln(1/p)\Bigr)}
 \\
 \nonumber
 \\
 \nonumber
  &\simeq&
 -\frac{1}{\pi} L^{2} \epsilon^{2} N
 \int_{|\tau|}^{1} \frac{dp}{p} \; \frac{1}{\Bigl(1 \; + \; \frac{2}{\pi}\, g_{0} \, \ln(1/p)\Bigr)}
 \\
 \nonumber
 \\
 &=&
 - \epsilon^{2} N \,C(\tau) ,
\label{68}
\end{eqnarray}
where
\begin{equation}
\label{69}
C(\tau) \; = \; \frac{1}{2 g_{0}} L^{2} \, \ln\Bigl(1 + \frac{2}{\pi}g_{0}\ln(1/|\tau|)\Bigr).
\end{equation}
Substituting this into the inverse Laplace transform relation (\ref{59}) we get:
\begin{equation}
  \label{70}
  {\cal P}_{\tau} (C) \; = \; \int_{-i\infty}^{+i\infty}\frac{ds}{2\pi i} \;
  \exp\Bigl\{  C(\tau) \, s \; - \;  C \, s  \Bigr\} \; = \;
  \delta\Bigl( C \, - \, C(\tau) \Bigr).
\end{equation}
This result shows that unlike the singular part of the internal energy the specific
heat in the vicinity of the critical point is a self-averaging quantity. In
particular, at large but finite value of the system size
$L \gg L_{*} \sim \exp(2/\pi g_{0})$ at the critical point at $\tau_{c} \sim 1/L$,
according to Eq.~(\ref{69}) the critical specific heat $C(L)$ scales with the system
size as
\begin{equation}
  \label{71}
  C(L)  \; \sim \; \frac{1}{2 g_{0}} L^{2} \, \ln\ln(L).
\end{equation}

Note, that the distribution (\ref{70}) describes only the singular
part of the specific heat, similar to the distributions (\ref{46})
and (\ref{53}) which describe the singular part of the internal
energy.  As a matter of fact, the singular part of the ``replica
free energy'' represented in Eq.~(\ref{68}) is linear in the replica
parameter $s=\epsilon^{2} N$.  Formally, by the inverse Laplace
transform this results in the $\delta$-function (\ref{70}), which
may be misleading as the specific heat of the system contains also a
regular part that is non-singular in the limit $\tau \to 0$. As we
have already mentioned, this last part is out of control for the
present renormalization group approach, however it is a random
quantity too. According to the central limit theorem this regular
part is normally distributed with its mean value proportional to the
volume of the system ($\sim L^{2}$, in the present case) and with a
variance proportional to the square-root of the volume of the system
($\sim \sqrt{L^{2}} = L$, in the present case). In other words, the
regular part of the specific heat can be represented as $C_{0}L^{2}
+ C_{*} L \zeta$ where the random variable $\zeta$ is normally
distributed with zero mean and unit variance, and the values of
$C_{0}$ and $C_{*}$ do not scale with $L$ as $L \to \infty$.
Correspondingly, in the replica representation this must give two
additional contributions to the replica free energy: in addition to
the expression presented in Eq.~(\ref{68}) one has two more terms
$C_{0}L^{2} s + (C_{*} L )^{2} s^2$ (where $s = \epsilon^{2} N$).
Thus, after the inverse Laplace transform the $\delta$-function
(\ref{70}) is replaced by a Gaussian distribution with the mean
value $C(\tau) + C_{0} L^{2}$, where $C(\tau)$ is given in
Eq.~(\ref{69})), and variance $C_{*} L$. In the limit $L \to \infty$
this results in the following behavior of the specific heat:
\begin{equation}
 C(\tau=1/L)  \sim L^{2}\ln\ln(L)   \gg  C_{0} L^{2} \gg  C_{*} L.
 \end{equation}
 Consequently, in this limit the distribution of the specific heat (which includes
 both the regular and the singular part) turns into a $\delta$-function centered at
 $C(L) \sim L^{2}\ln\ln(L)$, corresponding to self-averaging.

\section{Conclusions  \label{VI}}

We have derived an explicit expression for the probability distribution function of
the sample-to-sample fluctuations of the internal energy of the weakly disordered
critical two-dimensional Ising ferromagnet. The result obtained,
Eqs.~(\ref{52})--(\ref{53}), shows that the internal energy of this system is not
self-averaging. Instead, the typical value of its sample-to-sample fluctuations
scales in the same way as its average, proportional to $\sim L \ln\ln(L)$.  On the
other hand, the specific heat was shown here to exhibit self-averaging, with a
distribution function that converges to a $\delta$-function in the limit of infinite
system size. In contrast to the free energy of the system, which was discussed before
in Ref.~\cite{Dotsenko14}, the quantities discussed here are directly observable in
numerical simulations. It is not completely obvious at this point in how far the
singular behavior is masked in a lattice realization by the presence of regular
background terms and how clearly the lack or presence of self-averaging could be seen
experimentally. A numerical investigation of this system geared towards resolving
this intriguing question is the subject of a forthcoming study.

\acknowledgments

An essential part of this work was done during the {\em Ising lectures}, an annual
Workshop on Phase Transitions and Critical Phenomena held at the Institute for
Condensed Matter Physics, Lviv, Ukraine (May 17 - 19, 2016). V.D. is grateful to
Vladimir Dotsenko for numerous illuminating discussions. This work was supported in
part by the European Commission through the IRSES network DIONICOS under
  contract No. PIRSES-GA-2013-612707.

\appendix

\section{Derivation of Eq.~(\ref{41})}

\newcounter{A}
\setcounter{equation}{0}
\renewcommand{\theequation}{A.\arabic{equation}}

\vspace{5mm}

In this Appendix we explain in more details the derivation of Eq.~(\ref{41}). Let us
consider the quantity
\begin{equation}
\label{a1}
I(\tau) \; = \; \int_{|p|< 1} \frac{d^{2} p}{p^{2} + \tau^{2}}
\, f\bigl[\ln(1/p)\bigr],
\end{equation}
where $f(\xi)$ is a ``sufficiently good'' (not too divergent) function in the limit
$\xi\to\infty$, i.e., $\lim_{\xi\to\infty} f(\xi)\exp\{-\xi\} \to 0$. The leading
singularity of $I(\tau)$ in the limit $\tau \to 0$ can be estimated in the standard
way:
\begin{equation}
 \label{a2}
 I(\tau) \; \sim  \; \int_{|\tau| < |p|< 1} \frac{d^{2} p}{p^{2}}
\, f\bigl[\ln(1/p)\bigr]
\; \sim  \;
\int_{|\tau|}^{1} \frac{dp }{p} \, f\bigl[\ln(1/p)\bigr]
\; \sim \;
\int_{0}^{\ln(1/|\tau|)}  \, d\xi \; f(\xi),
\end{equation}
where $\xi = \ln(1/p)$. Now let us consider the slightly more complicated object
\begin{equation}
 \label{a3}
\tilde{I}(\tau) \; \equiv \; \int_{|p|< 1} \frac{d^{2} p}{p^{2} + m^{2}(p)}
\, f\bigl[\ln(1/p)\bigr],
\end{equation}
where instead of $\tau^{2}$ in the denominator we have a $p$-dependent mass term
$m^{2}(p)$,
\begin{equation}
 \label{a4}
m^{2}(p) \; = \; \frac{\tau^{2}}{1 + \frac{2}{\pi} g_{0} \ln(1/p)},
\end{equation}
which is the case when computing the specific heat singularity of the weakly
disordered 2D Ising model. One can consider two limiting cases:

\begin{enumerate}
\item[(a)] $\frac{2}{\pi} g_{0} \ln(1/|\tau|) \ll 1$ or $|\tau| \gg
  \exp(-\pi/2g_{0}) \equiv \tau_{*}$. In this case while
  integrating over $p$ one can just drop the presence of the
  nontrivial denominator in (\ref{a4}) and we get
  \begin{equation}
    \label{a5}
    \tilde{I}(\tau) \; \sim \; \int_{|\tau|}^{1} \frac{dp }{p} \,
    f\bigl[\ln(1/p)\bigr] \; \sim \; \int_{0}^{\ln(1/|\tau|)}  \, d\xi \;
    f(\xi),
  \end{equation}
  which coincides with the "pure" case (\ref{a2}).
\item[(b)]  $\frac{2}{\pi} g_{0} \ln(1/|\tau|) \gg 1$ or $|\tau| \ll
  \tau_{*}$. In this case we have
  \begin{equation}
    \label{a6}
    \tilde{I}(\tau) \; \equiv \; \int_{|p|< 1} \frac{d^{2} p}{p^{2} + m^{2}(p)}
    \, f\bigl[\ln(1/p)\bigr] \; \sim \; \int_{p_{*}(\tau)}^{1}
    \frac{dp}{p} \, f\bigl[\ln(1/p)\bigr],
  \end{equation}
  where $p_{*}(\tau)$ is defined by the condition:
  \begin{equation}
    \label{a7}
    p_{*} \; \sim \; \frac{|\tau|}{\sqrt{g_{0}\ln(1/p_{*})}},
  \end{equation}
  which yields
  \begin{equation}
    \label{a8}
    p_{*}(\tau) \; \sim \;
    \frac{|\tau|}{\sqrt{g_{0}\ln(1/|\tau|)}}.
  \end{equation}
  Substituting this in (\ref{a6}), we get
  \begin{equation}
    \label{a9}
    \tilde{I}(\tau) \; \sim \; \int_{0}^{\xi_{*}(\tau)} \; d\xi \; f(\xi),
  \end{equation}
  where in the limit $|\tau| \to 0$,
  \begin{equation}
    \label{a10}
    \xi_{*}(\tau) \; \sim \;
    \ln\Bigl[\frac{\sqrt{g_{0}\ln(1/|\tau|)}}{|\tau|}\Bigr] \; = \;
    \ln(1/|\tau|) \; + \; \frac{1}{2}\ln\ln(1/|\tau|) \; + \; \frac{1}{2}\ln(g_{0})  \; \sim \;
    \ln(1/|\tau|) \; + \; O\Bigl(\ln\ln(1/|\tau|)\Bigr).
  \end{equation}
  Thus, in this case we get
  \begin{equation}
    \label{a11}
    \tilde{I}(\tau) \; \sim \; \int_{0}^{\ln(1/|\tau|)}  \, d\xi \; f(\xi),
  \end{equation}
  which means that in the limit $\tau \to 0$ in both cases (a) and (b) we can cut the
  integration over $p$ at $p_{*} \sim \tau$. Note that in the considered model
  $|\tau|^{-1} \sim R_{c}$ is the correlation of the pure system, and the presence of
  disorder produces not more than a logarithmic correction to the correlation length.
\end{enumerate}
In the case considered in this paper the situation is somewhat more tricky due to the
presence of the second term in the brackets of Eq.~(\ref{35}). According to
Eqs.~(\ref{35})--(\ref{37}) in the limit $\tau \to 0$ (at $|\tau| \ll \tau_{*}$ where
$g_{0}\xi \gg 1$) we have
\begin{equation}
 \label{a12}
 m(p) \; \sim \;  \frac{\tau \; + \; \frac{1}{2} (\epsilon N) \, \ln\Bigl( g_{0}
   \ln(1/p)\Bigr)}{\sqrt{g_{0}\ln(1/p)}}.
\end{equation}
According to the standard logic of the replica technique, first we have to assume
that for a given value of the replica parameter $N$ the value of the parameter
$\epsilon$ is considered to be less than everything else, such that
$(\epsilon N) \, \ln\ln(1/|\tau|) \ll 1$, so that the integration over $p$ is cut at
$p_{*} \sim \tau$ as in the above examples.  On the other hand, in the further
inverse Laplace transform integration over analytically continued complex parameter
$N$, its relevant value turns out to be of order $1/\epsilon$, which means that the
relevant value of the (complex) product $(\epsilon N)$ turns out to be finite.

\end{document}